\documentclass[aps,pra,twocolumn,tightlines,superscriptaddress]{revtex4-1}  % for review and submission

\usepackage{graphicx}  % needed for figures
\usepackage{dcolumn}   % needed for some tables
\usepackage{pbox}
\usepackage{bm}        % for math
\usepackage{amsmath,amsfonts,amssymb}
\usepackage{url,hyperref}

\newcommand{\sref}[1]{Sec. \ref{#1}}
\newcommand{\eref}[1]{Eq.~(\ref{#1})}
\newcommand{\tref}[1]{Table~\ref{#1}}

\begin{document}

\title{Polarizabilities of low-lying states of silver}

\author{S.~G.~Porsev}
\affiliation{Department of Physics and Astronomy, University of Delaware, Newark, Delaware 19716, USA}
\author{D.~Filin}
\affiliation{Department of Physics and Astronomy, University of Delaware, Newark, Delaware 19716, USA}
\author{C.~Cheung}
\affiliation{Department of Physics and Astronomy, University of Delaware, Newark, Delaware 19716, USA}
\author{M.~S.~Safronova}
\affiliation{Department of Physics and Astronomy, University of Delaware, Newark, Delaware 19716, USA}

\begin{abstract}
Assembly of ultracold polar molecules containing silver (Ag)  from laser-cooled atoms requires knowledge of the dynamic polarizabilities of Ag at convenient laser wavelengths. 
We present calculations and analysis of the energies and electric-dipole dc and ac polarizabilities of the low-lying states of neutral Ag. Calculations of the properties of the $4d^{10}x$ states, where $x=5s,6s,7s,5p,6p,7p,5d,6d$, and $4f$, are performed using the linearized coupled cluster single-double method.  The properties of the $4d^9 5s^2\,^2\!D_{5/2,3/2}$ states are obtained within the framework of configuration interaction with 11 and 17 electrons in the valence field. We analyze the different contributions to the polarizabilities and estimate the uncertainties of our predictions. 
\end{abstract}

\date{\today}

\maketitle
%------------------------------
\section{Introduction}
%------------------------------
The transition-metal silver atom has recently begun to attract increased attention from theorists and experimentalists.
The prospect of using the electric quadrupole $4d^{10} 5s\,\, ^2\!S_{1/2} -\, 4d^9 5s^2\,\,^2\!D_{5/2}$ transition as a
transition in the optical lattice clock was studied in Ref.~\cite{DzuAllFla21}. This very narrow transition was observed in
Ref.~\cite{BadPliJun06} and the hyperfine transition frequencies in the 107 and 109 isotopes of Ag were measured using two-photon laser spectroscopy.

One of the interesting features of silver is its ability to form ultracold, highly polar diatomic molecules containing the silver atom (in its ground state) interacting with a noble gas~\cite{JouLarMar91,BroDun95,SunAbeHad19}, an alkali metal, or an alkaline earth metal atom~\cite{SmiTom21}.
Ag-alkaline-earth-metal molecules were predicted to have exceptionally large dipole moments that exceeded those typically observed in alkali-metal dimers \cite{SmiTom21}, motivating the use of Ag-based molecules for quantum simulations, ultra-cold chemistry, and fundamental physics.
The possible sensitivity of the diatomic molecule AgPb to the electron electric dipole moment was suggested in a recent work~\cite{StuRicChe24}.
The RaAg molecule was proposed to probe new physics beyond the standard model and to search
for the electric dipole moment of the electron and the scalar-pseudoscalar interaction \cite{SunAbeHad19}.
Experimental efforts toward next-generation EDM searches with molecules containing Ag are underway \cite{DeMille24}.

To address questions in condensed matter physics and quantum dynamics by achieving full quantum control over all degrees
of freedom in a molecular gas, the potassium-silver molecule (KAg) was proposed for a study by the University of Chicago
group~\cite{Yan24}.
%A special property of this molecule is that the electrons are highly unevenly distributed, forming an ``electric dipole moment.'' 
Compared to other ultra-cold molecules, the electric dipolar interaction of KAg
is expected to be an order of magnitude stronger \cite{SmiTom21}, facilitating engineering and detecting many-body effects arising from interactions~\cite{Yan24}.

However, proposed experiments require the assembly of ultracold polar molecules containing silver from laser-cooled atoms, which requires knowledge of the dynamic polarizabilities of Ag at convenient laser wavelengths.
To support the experimental efforts, we calculated the dc and ac polarizabilities at the 532~nm and 1064~nm wavelengths, convenient for laser trapping,
for the relevant states of Ag and evaluated their uncertainties. 

A specific feature of the Ag atom is the presence of low-lying states with the unfilled $4d$ shell, $4d^9 5s^2\,\, ^2\!D_{3/2,5/2}$, 
along with the states belonging to the $4d^{10} x$ configuration (where $x \equiv 5,6s; 5,6p, 5d$, etc.), complicating the accurate prediction of the atomic properties of Ag. 

To calculate the properties of the $4d^{10} x$ states, we consider Ag as an atomic system with a single valence electron above the
core $[1s^2, 2s^2,...,4d^{10}]$. We need to accurately take into account the correlations between the valence and core electrons. We perform computations using the all-order linearized coupled cluster single-double (LCCSD) method. To evaluate an uncertainty, we also carry out the computations using many-body perturbation theory (MBPT) over the
residual Coulomb interaction.

For calculating the properties of the $4d^9 5s^2\,\, ^2\!D_{3/2,5/2}$ states, a single electron approach is not applicable.
In this case, we consider Ag as an atom with many valence electrons and apply the configuration
interaction (CI) method.  11- and 17-electron (17$e$) CI calculations are carried out, assuming that (i) $4d$ and (ii) $4d$ and $4p$
electrons are in the valence field. In the following, we describe the computations and discuss the results.
%----------------------------------
\section{Single-electron approach}
\label{Sing_e}
%----------------------------------
We consider Ag as a univalent atom with a core $[1s^2, 2s^2,...,4d^{10}]$ and a valence electron above it.
The initial Dirac-Hartree-Fock (DHF) self-consistency procedure included the Breit interaction and was performed for the core
electrons. Then, the $5$--$7s$, $5$--$7p$, $5d$, and $4f$ orbitals were constructed in the frozen core potential.
The remaining virtual orbitals were formed using a recurrent procedure described in Refs.~\cite{KozPorFla96,KozPorSaf15}.
The newly constructed functions were then orthonormalized with respect to the functions of the same symmetry.
The basis sets included a total of six partial waves ($l_{\rm max} = 5$) and orbitals with a principal quantum number $n$ of up to 35.
%The Breit interaction was included in the stage of constructing the basis set.

In our approach, the wave functions and energy levels of the valence electrons were found by solving the
relativistic equation~\cite{DzuFlaKoz96},
\begin{equation}
H(E_n) \Phi_n = E_n \Phi_n,
\label{Heff}
\end{equation}
where the effective Hamiltonian is defined as
\begin{equation}
H(E) = H_{\rm FC} + \Sigma(E).
\label{Heff1}
\end{equation}
Here $H_{\rm FC}$ is the Hamiltonian in the frozen-core approximation, and
the energy-dependent operator $\Sigma(E)$, accounting for the virtual excitations of the core electrons,
was constructed in two ways: using (i) the second-order MBPT over the residual Coulomb interaction~\cite{DzuFlaKoz96}
and (ii) the linearized coupled cluster single-double (all-order) method~\cite{SafKozJoh09}.
In the following, we refer to these approaches as MBPT and all-order methods. The difference between the results obtained by
these two methods allows us to estimate the uncertainty of our calculation.
%----------------------
\subsection{Energy levels}
%----------------------
We started by calculating the low-lying energy levels. The results are presented in \tref{Tab:E}.
The lowest order DHF contribution to the energies is labeled ``DHF.'' The results obtained in the framework of the MBPT and
all-order methods are given in the rows labeled ``MBPT'' and ``All'', respectively.
For the ground state, we present its removal energy, which can be compared to the ionization potential
IP(Ag$^+$)~\cite{RalKraRea11}. For the excited states, the excitation energies are displayed.
The experimental values from the NIST database~\cite{RalKraRea11} are given in the column labeled ``Exp.''
The difference between the experimental and theoretical ``DHF,'' ``MBPT,'' and ``All'' values is presented in the
last three columns. As follows from the table, the difference between the experimental energies and those
obtained in the framework of the all-order method is overall better than 1\%. The remaining difference can be
attributed to a contribution of triple excitations, quantum electrodynamical corrections, and corrections from
the higher partial waves.
%Such an accuracy is sufficient for our purposes.
% ##########################################################################################
\begin{table*}[htp]
\caption{The energies of the low-lying states, calculated in the DHF (labeled as ``DHF''), MBPT (labeled as ''MBPT''),
and all-order (labeled as ``All'') approximations are presented. 
For the ground state, we present its removal energy (in cm$^{-1}$) which can be compared to the ionization potential, 
IP(Ag$^+$)~\cite{RalKraRea11}. For the excited states, the excitation energies (in cm$^{-1}$) are displayed.
The experimental values from the NIST database~\cite{RalKraRea11} are given in the column labeled ``Exp.''
The difference between the experiment and theory is presented in the last three columns.}
\label{Tab:E}%
\begin{ruledtabular}
\begin{tabular}{lccccccc}
                     &   DHF    &   MBPT   &   All   &Exp.~\cite{RalKraRea11}& Exp.-HFD & Exp.-MBPT  & Exp.-All\\
\hline \\[-0.6pc]
$5s\,\, ^2\!S_{1/2}$ &  50337   &  61991   &  61295  &  61106                &    18\%  &    -1\%     & -0.3\% \\[0.2pc]
$6s\,\, ^2\!S_{1/2}$ &  33228   &  43490   &  42885  &  42556                &    22\%  &    -2\%     & -0.8\% \\[0.2pc]
$5d\,\, ^2\!D_{3/2}$ &  38354   &  49664   &  48946  &  48744                &    21\%  &    -2\%     & -0.4\% \\[0.2pc]
$5d\,\, ^2\!D_{5/2}$ &  38369   &  49685   &  48965  &  48764                &    21\%  &    -2\%     & -0.4\% \\[0.2pc]
$7s\,\, ^2\!S_{1/2}$ &  41684   &  52848   &  52126  &  51887                &    20\%  &    -2\%     & -0.5\% \\[0.2pc]
$6d\,\, ^2\!D_{3/2}$ &  43778   &  55261   &  54566  &  54203                &    19\%  &    -2\%     & -0.7\% \\[0.2pc]
$6d\,\, ^2\!D_{5/2}$ &  43788   &  55273   &  54578  &  54214                &    19\%  &    -2\%     & -0.7\% \\[0.2pc]
%$8s\,\, ^2\!S_{1/2}$ &  47153   &  58262   &  57546  &  55581                &    15\%  &    -5\%     & -3.5\% \\[0.5pc]
\hline  \\[-0.6pc]
$5p\,\, ^2\!P_{1/2}$ &  23628   &  30550   &  29809  &  29552                &    20\%  &    -3\%     & -0.9\% \\[0.2pc]
$5p\,\, ^2\!P_{3/2}$ &  24202   &  31512   &  30728  &  30473                &    21\%  &    -3\%     & -0.8\% \\[0.2pc]
$6p\,\, ^2\!P_{1/2}$ &  38557   &  49281   &  48594  &  48297                &    20\%  &    -2\%     & -0.6\% \\[0.2pc]
$6p\,\, ^2\!P_{3/2}$ &  38722   &  49496   &  48802  &  48501                &    20\%  &    -2\%     & -0.6\% \\[0.2pc]
$7p\,\, ^2\!P_{1/2}$ &  43830   &  55187   &  54488  &  54041                &    19\%  &    -2\%     & -0.8\% \\[0.2pc]
$7p\,\, ^2\!P_{3/2}$ &  43887   &  55289   &  54593  &  54121                &    19\%  &    -2\%     & -0.9\% \\[0.2pc]
$4f\,\, ^2\!F_{5/2}$ &  44000   &  55433   &  54737  &  54205                &    19\%  &    -2\%     & -1.0\% \\[0.2pc]
$4f\,\, ^2\!F_{7/2}$ &  44088   &  55489   &  54793  &  54205                &    19\%  &    -2\%     & -1.1\%
\end{tabular}
\end{ruledtabular}
\end{table*}
% ##########################################################################################
%----------------------------
\subsection{Polarizabilities}
%----------------------------
We find the static and dynamic electric-dipole ($E1$) polarizabilities for the lowest-lying even- and
odd-parity states of Ag at the specific wavelengths 532~nm and 1064~nm. The expression for an $E1$ ac polarizability at the frequency
$\omega$ of the state $|JM\rangle$ (where $J$ is the total angular momentum and $M$ is its projection) can be written (in a.u.) as
%--------------------------------------------------------------------
\begin{equation}
\alpha(\omega) = 2 \sum_n \frac{(E_n-E) |\langle JM|d_z|n\rangle|^2}{(E_n-E)^2 - \omega^2} ,
\label{alpha}
\end{equation}
%--------------------------------------------------------------------
where $\bf d$ is an electric dipole moment operator and $E$ and $E_n$ are the energies of the initial and intermediate states, respectively.

The expression for $\alpha$ can be rewritten in the following form:
%--------------------------------------------------------------------
\begin{eqnarray}
\alpha(\omega) &=& \sum_n \langle JM|d_z|n\rangle \,\langle n|d_z| JM\rangle \nonumber \\
       &\times& \left[ \frac{1}{E_n-E+\omega} + \frac{1}{E_n-E-\omega} \right] .
\label{alpha1}
\end{eqnarray}
%--------------------------------------------------------------------

We consider Ag as an atom with one valence electron above the closed core.
The polarizability can be written as
\[
\alpha \equiv \alpha_v + \alpha_c %+ \alpha_{vc}
\]
where $\alpha_v$ and $\alpha_c$ are the valence and core contributions.

To find $\alpha_v$, we use the Sternheimer~\cite{Ste50} or Dalgarno-Lewis~\cite{DalLew55} method and solve inhomogeneous equations
%------------------------------------------------------------------
\begin{eqnarray}
(H - E \pm \omega)\, |\delta \phi_{\pm} \rangle = d_z\, |\Phi \rangle,
\label{inhom}
\end{eqnarray}
%------------------------------------------------------------------
where $\Phi$ is an eigenstate of the Hamiltonian $H$.
The wave function $|\delta \phi_{\pm} \rangle$ can be found from \eref{inhom} as
%------------------------------------------------------------------
\begin{eqnarray}
|\delta \phi_{\pm} \rangle &=& \frac{1}{H - E \pm \omega} d_z |\Phi \rangle \nonumber \\
&=& \sum_n \frac{1}{H - E \pm \omega}|n\rangle \langle n |d_z |\Phi \rangle,
\label{phi}
\end{eqnarray}
%------------------------------------------------------------------
where we use the closure relation $\sum_n | n \rangle \langle n | = 1$.

Then, using~\eref{alpha1}, we can find the valence contribution $\alpha_v$ as
\begin{eqnarray}
\alpha_v(\omega) =  \langle \Phi| d_z |\delta \phi_{+} \rangle + \langle \Phi| d_z |\delta \phi_{-} \rangle .
\label{alpha2}
\end{eqnarray}

The core contribution $\alpha_c$ is calculated in the single-electron approximation using
a sum-over-state approach. The single-electron matrix elements (MEs) of the electric dipole operator include the random-phase approximation (RPA)
corrections. Note that in calculating $\alpha_c$, a core electron can be excited to the occupied valence state.
The Pauli principle forbids this. We take this into account by subtracting this contribution from $\alpha_c$.

Disregarding the vector polarizability, we can present the expression $\alpha(\omega)$ as the sum of the scalar and tensor parts,
\begin{equation}
\alpha(\omega) = \alpha_0 + \alpha_2 \, \frac{3M^2-J(J+1)}{J(2J-1)} .
\label{al_st}
\end{equation}

To determine uncertainties of the polarizabilities, we calculated them in three ways. The first two are the
MBPT and all-order methods,  where we include only RPA corrections to the electric dipole operator.
The third and most complete calculation additionally includes the smaller corrections to the operator $\bf d$ beyond RPA, such as
the core-Brueckner, structural radiation, and normalization corrections (see Refs.~\cite{DzuKozPor98,DzuFlaKoz96} for details).
We designate this approximation as all-order+AC, where the abbreviation ``AC'' means all corrections.

The results obtained in these approximations for the scalar parts of the polarizabilities are presented in~\tref{Tab:Polar0}.
In most cases, there are several low-lying intermediate states (see~\eref{alpha}) that give a dominant contribution to
the polarizability. The final values for such polarizabilities are obtained by replacing theoretical energies with experimental
ones in the dominant contributions.
The only exclusions are the dynamic polarizabilities of the $6p\,^2\!P_{1/2,3/2}$ states, for which
the contribution of high-lying states is substantial. For these polarizabilities, we did not make such a replacement.

To assign uncertainties to the polarizabilities, we need to take into account the uncertainties of the valence
and core parts, $\alpha_v$ and $\alpha_c$. 

The former were determined for most polarizabilities based on the difference
between the MBPT and all-order+AC results.
In two cases, the uncertainties were determined in a different way.
As seen from \tref{Tab:Polar0}, the scalar polarizabilities of the $5d\,^2\!D_{3/2,5/2}$ states
calculated at $\lambda = 1064$ nm are very insensitive to the high-order corrections to the wave functions and corrections
to the electric dipole operator. However, comparing the final and all-order+AC results, we see that they are sensitive to replacement
of the theoretical energies with the experimental ones.
In these particular cases, the uncertainty was determined as half of the difference between the final and all-order+AC values.

Another source of uncertainty is the core polarizability. They were calculated in the single-electron approximation. This method is not very accurate, and we assume that the uncertainty of $\alpha_c$
is about 10\%. The core part of the polarizability is rather insensitive to the frequency and is virtually the
same for the static and dynamic polarizabilities. We find it to be $\alpha_c = 8.8 (0.9)\,\, {\rm a.u.}$ for the even and
$6p\,\, ^2\!P_J$ states. For the $5p\,\, ^2\!P_J$ states, $\alpha_c = 8.4 (0.9)\,\, {\rm a.u.}$.
% ###################################################################################################################
\begin{table}[htp]
\caption{The dc and ac ($\lambda = 532$ and 1064 nm) scalar polarizabilities $\alpha_0$ (in a.u.) of the low-lying states, calculated in the MBPT, all-order (labeled as ``All''), and all-order+AC
(labeled as ``All+AC'') approximations, are presented. The final (recommended) values are given in the column
labeled ``Final.'' The uncertainties are given in parentheses.}
\label{Tab:Polar0}%
\begin{ruledtabular}
\begin{tabular}{lccccc}
         &                     &   MBPT   &   All     & All+AC &         Final       \\
\hline \\ [-0.6pc]
$5s\,\, ^2\!S_{1/2}$ &  Static &   48.3   &   50.2    &   49.1   & $     49.5(1.2)$    \\[0.2pc]
                     &  532 nm &   70.9   &   75.8    &   74.0   & $     75.3(3.3)$    \\[0.2pc]
                     & 1064 nm &   52.3   &   54.6    &   53.4   & $     53.8(1.4)$    \\[0.5pc]
$6s\,\, ^2\!S_{1/2}$ &  Static &    1768  &   1822    &   1816   & $     1805(47)$     \\[0.2pc]
                     &  532 nm &    -111  &   -105    &   -106   & $     -108(5)$      \\[0.2pc]
                     & 1064 nm &   -1579  &  -1525    &  -1514   & $    -1533(64)$     \\[0.5pc]
$5d\,\, ^2\!D_{3/2}$ &  Static &  -17540  & -19628    & -19556   & $   -13600(2000)$   \\[0.2pc]
                     &  532 nm &   -1210  &  -1124    &  -1109   & $     -935(100)$    \\[0.2pc]
                     & 1064 nm &    -921  &   -926    &   -921   & $     -795(65)$     \\[0.5pc]
$5d\,\, ^2\!D_{5/2}$ &  Static &  -30472  & -35312    & -35186   & $   -21100(4700)$   \\[0.2pc]
                     &  532 nm &     583  &    673    &    662   & $      770(80)$     \\[0.2pc]
                     & 1064 nm &   -1001  &  -1006    &  -1002   & $     -853(75)$     \\[0.5pc]
$5p\,\, ^2\!P_{1/2}$ &  Static &     142  &    141    &    139   & $      139(3)$      \\[0.2pc]
                     &  532 nm &    2310  &   2163    &   2132   & $     1840(180)$    \\[0.2pc]
                     & 1064 nm &     219  &    214    &    212   & $      213(7)$      \\[0.5pc]
$5p\,\, ^2\!P_{3/2}$ &  Static &     167  &    165    &    163   & $       103(18)$     \\[0.2pc]
                     &  532 nm &   -1257  &  -1412    &  -1392   & $     -1540(135)$   \\[0.2pc]
                     & 1064 nm &     292  &    280    &    277   & $       279(16)$    \\[0.5pc]
$6p\,\, ^2\!P_{1/2}$ &  Static &   28025  &  30262    &  30158   & $     24035(2130)$  \\[0.2pc]
                     &  532 nm &    -765  &   -743    &   -742   & $      -740(25)$    \\[0.2pc]
                     & 1064 nm &    -127  &   -101    &   -100   & $      -100$        \\[0.5pc]
$6p\,\, ^2\!P_{3/2}$ &  Static &   56178  &  64363    &  64141   & $     40400(8000)$  \\[0.2pc]
                     &  532 nm &    -463  &   -464    &   -464   & $      -465$        \\[0.2pc]
                     & 1064 nm &      39  &     48    &     50   & $        50$
\end{tabular}
\end{ruledtabular}
\end{table}
% ####################################################################################################################

For the dc and ac polarizabilities of the ground state, the uncertainties of $\alpha_v$ and $\alpha_c$ are comparable. Our final value for the static scalar polarizability of the ground state
$\alpha_0(5s\,\, ^2\!S_{1/2}) = 49.5(1.2)$ a.u. is in good agreement with the recommended value $55(8)$ a.u.~\cite{SchNag19} obtained by compiling theoretical and experimental results.
For all other states, the uncertainty of $\alpha_v$ gives a dominant contribution to the uncertainty budget.

Calculating the static scalar and tensor polarizabilities of the $4d^{10} 5p\,\, ^2\!P_{3/2}$ state takes some care. The even state $4d^9 5s^2\,\, ^2\!D_{5/2}$ is separated from $4d^{10} 5p\,\, ^2\!P_{3/2}$ by the small energy interval 230 cm$^{-1}$ and can contribute to the polarizability of $^2\!P_{3/2}$. The properties of states with the unfilled $4d$ shell cannot be studied in the framework of a single-electron approach. To do that, we apply the CI method. We will discuss in detail the calculation performed within the framework of the CI method in Section~\ref{CI}. Here we mention only the main results. Using the CI method, we obtain
$|\langle 4d^9 5s^2\,\, ^2\!D_{5/2} || d || 5p\, ^2\!P_{3/2} \rangle| = 0.61(9)\,\, {\rm a.u.}$ and the transition rate $W (^2\!P_{3/2} \rightarrow ^2\!D_{5/2}) \approx 2.3 \,\, {\rm s}$ that can be compared with the experimental value $1.6(6)\,{\rm s}$~\cite{UhlDirWal00}. The experimental result is not very precise and does not allow us to accurately determine the theoretical uncertainty. Based on the difference of 30\% between our result and the experimental central value, we estimate the uncertainty of ME at the level of 15\%.

Using this ME and experimental energy levels, we can easily calculate the contribution of the $4d^9 5s^2\,\, ^2\!D_{5/2}$ state to the static scalar and tensor polarizabilities of the $5p\, ^2\!P_{3/2}$ state to be $-60(18)$ a.u. and $12(4)$ a.u., respectively. These contributions were taken into account in the final values of $\alpha_0$ and $\alpha_2$ presented in Tables~\ref{Tab:Polar0} and~\ref{Tab:Polar2}.
We note that for all other dc and ac polarizabilities of odd states, the contribution of $4d^9 5s^2\,\, ^2\!D_{3/2,5/2}$ is small and is within the assigned uncertainties.

For $\alpha_0(6p\,^2\!P_{1/2})$ at $\lambda = 1064$ nm and
$\alpha_0(6p\,^2\!P_{3/2})$ at $\lambda = 532$ and $1064\, {\rm nm}$, the uncertainties are not assigned.
This is due to large contributions from high-lying states that are difficult to control and large cancellations
between different contributions. For example, when the two MEs in \eref{alpha2} are close to each other in
absolute value but are of opposite sign, they substantially cancel each other out, significantly worsening the accuracy of the final value.

In \tref{Tab:Polar2}, we present the tensor polarizabilities for states with total angular momentum $J > 1/2$.
The final values of the tensor polarizabilities and their uncertainties were determined in the same way as was done for the scalar
polarizabilities. Since there is no core contribution to tensor polarizability, the valence part determines its value and uncertainty. The designations used in the table are the same as in~\tref{Tab:Polar0}.
% ###################################################################################################################
\begin{table}[htp]
\caption{The dc and ac tensor polarizabilities $\alpha_2$ (in a.u.) of the low-lying states,
calculated in the MBPT, all-order (labeled ``All'') and all-order+AC (labeled ``All+AC'') approximations, are presented.
The final (recommended) values are given in the column labeled ``Final.'' The uncertainties are given in parentheses.}
\label{Tab:Polar2}%
\begin{ruledtabular}
\begin{tabular}{lccccc}
         &                     &  MBPT     &   All     &  All+AC  &      Final      \\
\hline \\ [-0.6pc]
$5d\,\, ^2\!D_{3/2}$ &  Static &   8140    &   8499    &   8468   & $   7650(330)$  \\[0.2pc]
                     &  532 nm &   1290    &   1230    &   1213   & $   1073(77)$   \\[0.2pc]
                     & 1064 nm &    203    &    205    &    205   & $    175(15)$   \\[0.5pc]
$5d\,\, ^2\!D_{5/2}$ &  Static &  31755    &  36584    &  36457   & $  22420(4700)$ \\[0.2pc]
                     &  532 nm &   -709    &   -800    &   -788   & $   -885(80)$   \\[0.2pc]
                     & 1064 nm &    341    &    343    &    342   & $    291(26)$   \\[0.5pc]
$5p\,\, ^2\!P_{3/2}$ &  Static &    -58    &    -55    &    -55   & $    -43(5)$    \\[0.2pc]
                     &  532 nm &    164    &    191    &    188   & $    204(15)$   \\[0.2pc]
                     & 1064 nm &   -174    &   -139    &   -137   & $   -140(13)$   \\[0.5pc]
$6p\,\, ^2\!P_{3/2}$ &  Static &  -5725    &  -6396    &  -6396   & $  -4420(650)$  \\[0.2pc]
                     &  532 nm &     39    &     38    &     38   & $     38$       \\[0.2pc]
                     & 1064 nm &   -106    &   -112    &   -111   & $   -111$
\end{tabular}
\end{ruledtabular}
\end{table}
% ####################################################################################################################

%-----------------------------------------
\section{Configuration interaction method}
\label{CI}
%-----------------------------------------
As seen in the NIST database~\cite{RalKraRea11}, there are two low-lying states with the unfilled $4d$ shell
($4d^9 5s^2\,\, ^2\!D_{3/2,5/2}$) whose properties are of interest to experimentalists. These properties
cannot be studied in the framework of a single-electron method, so the configuration interaction method is used instead. Here, we utilize the pCI software package for computations~\cite{2025pCI}.

We constructed the basis set in a different way than in the single-electron approach. The initial self-consistency DHF procedure was performed for
the $4d^9 5s^2$ configuration. Then, all electrons were frozen and an electron was moved from the $5s$ to $5p$ shell, to construct
the $5p_{1/2,3/2}$ orbitals for the $4d^9 5s5p$ configuration. Other DHF orbitals
were constructed for the $4d^{10} x$ configurations, where $x \equiv 4f,5d,6s,6p,7s,7p$.
The remaining virtual orbitals were formed using a recurrent procedure described in Refs.~\cite{KozPorFla96,KozPorSaf15}.
In total, the basis set included five partial waves ($l_{\rm max} = 4$) and orbitals with the principal quantum number $n$ up to 25.
\subsection{Energies}
%--------------------
We carried out CI calculations with 11 and 17 electrons in the valence field. In the first case, we include the $4d$ electrons in the valence field, doing the calculation within the framework of the 11-electron (11$e$) CI. In the second case, we include the $4d$ and
$4p$ electrons in the valence field, performing the 17$e$ CI calculation.
For the 11$e$ CI, the set of configurations was constructed by including single and double excitations
from the main configurations, $4d^{10}(5s,6s,7s)$ and $4d^9(5s^2,5p^2)$ for the even-parity states and
$4d^{10}(5p,6p,7p)$ and $4d^9(5s5p,5s6p)$ for the odd-parity states, to the shells up to $12s,12p,12d,12f$
(we designate it as $[12spdf]$). For the 17$e$ CI, the main configurations remained the same, but single and
double excitations were also allowed from the $4p$ shell.

Since our goal is to calculate the polarizabilities of the $4d^9 5s^2\, ^2\!D_{5/2,3/2}$ states, our main focus is the odd-parity
states that can contribute a lot to these polarizabilities. From general
considerations, we can expect a large contribution from the states belonging to the configuration $4d^9 5s5p$ because there is a single-electron electric-dipole $5p - 5s$ transition between these configurations.
Furthermore, strong electric-dipole transitions can be expected from $^2\!D_{5/2,3/2}$ to odd states with the same total
spin $S=1/2$. As seen in the NIST database~\cite{RalKraRea11}, there are such odd states, but they lie very high
(above the ionization limit), making their accurate calculation particularly difficult.

In~\tref{E:CI}, we present the energies of the $4d^9 5s^2\, ^2\!D_J$ states and the odd states, giving a large contribution to the $4d^9 5s^2\, ^2\!D_J$ polarizabilities obtained in the framework of the 11$e$ and 17$e$ CI methods. To test the sensitivity of these energies 
%(and subsequently $4d^9 5s^2\, ^2\!D_J$ polarizabilities)
to the method of constructing the basis set and size of the CI space, we performed another CI calculation.
We used the basis set constructed in $V^{N-1}$ approximation (which we applied in the single-electron approach and described
in \sref{Sing_e}) and allowed single, double, and some triple excitations to $[20s19pdfg]$. In this way, the CI space
was substantially extended. These results are labeled ``$[20s19pdfg] $11$e$ CI'' in~\tref{E:CI}.

For calculating polarizabilities, we need to have the correct energy difference between the $4d^9 5s^2\, ^2\!D_J$ state and an odd-parity
state. To follow it, the excitation energies, presented in~\tref{E:CI}, are counted from the $4d^9 5s^2\, ^2\!D_{5/2}$ state.
% ##########################################################################################
\begin{table}[htp]
\caption{The energies (in cm$^{-1}$) of the even- and odd-parity levels calculated in the framework of the 11$e$ and
17$e$ CI methods with excitations to $[12spdf]$ and the 11$e$ CI method with excitations to $[20s19pdfg]$ are presented.
The excitation energies are counted from the $4d^9 5s^2\, ^2\!D_{5/2}$ state.
The experimental values from the NIST database~\cite{RalKraRea11} are given in the last column.}
\label{E:CI}%
\begin{ruledtabular}
\begin{tabular}{lcccc}
\multicolumn{1}{c}{} & \multicolumn{2}{c}{$[12spdf]$} & \multicolumn{1}{c}{$[20s19pdfg]$} & \multicolumn{1}{c}{} \\
                             & 11$e$ CI & 17$e$ CI & 11$e$ CI & Experiment \\
\hline \\[-0.6pc]
$4d^9 5s^2\,\, ^2\!D_{5/2}$  &    0     &    0     &     0     &    0    \\[0.2pc]
$4d^9 5s^2\,\, ^2\!D_{3/2}$  &  4325    &  4450    &    4368      &  4472   \\[0.2pc]
 \\ [-0.6pc]
%$4d^{10} 5p\,\, ^2\!P_{1/2}$ &  -320    &  1806    &          &   -690  \\[0.2pc]
$4d^{10} 5p\,\, ^2\!P_{3/2}$ &   394    &  2520    &   -754   &    230  \\[0.3pc]

$4d^9 5s5p\,\, ^2\!P_{3/2}$  &  42067   &  42664   &    44680 &  41942  \\[0.2pc]
$4d^9 5s5p\,\, ^2\!F_{7/2}$  &  42393   &  42937   &    44900 &  42092  \\[0.2pc]
$4d^9 5s5p\,\, ^2\!D_{5/2}$  &  43950   &  44566   &    46156 &  43285  \\[0.2pc]
$4d^9 5s5p\,\, ^2\!P_{1/2}$  &  46260   &  47008   &    48976 &  46162  \\[0.2pc]
$4d^9 5s5p\,\, ^2\!F_{5/2}$  &  46838   &  47517   &    49345 &  46568  \\[0.2pc]
$4d^9 5s5p\,\, ^2\!D_{3/2}$  &  48293   &  49038   &    50527 &  47700
\end{tabular}
\end{ruledtabular}
\end{table}
% ##########################################################################################
Comparing the theoretical and experimental results, we see that the largest difference does not exceed 3\%
for the high-lying states. Such an accuracy is sufficient for our purposes.
%----------------------------
\subsection{Polarizabilities}
%----------------------------
To calculate the polarizabilities of the $4d^9 5s^2\, ^2\!D_{5/2,3/2}$ states, we again used the method of solution of the
inhomogeneous equation described in Section II.B. The results of the calculation of the static and dynamic scalar
and tensor polarizabilities of the $4d^9 5s^2\,\, ^2\!D_{5/2,3/2}$ states are presented in~\tref{Tab:PolarCI}.
The final (recommended) values are given in the column labeled ``Final.''
% ###################################################################################################################
\begin{table}[t]
\caption{The static and dynamic scalar ($\alpha_0$) and tensor ($\alpha_2$) polarizabilities (in a.u.) of
the $4d^9 5s^2\, ^2\!D_{5/2,3/2}$ states, calculated in the framework of the CI method, are presented. The final (recommended) values
are given in the column labeled ``Final.''}
\label{Tab:PolarCI}%
\begin{ruledtabular}
\begin{tabular}{lcccccc}
\multicolumn{1}{c}{} & \multicolumn{1}{c}{} & \multicolumn{1}{c}{} &
\multicolumn{2}{c}{$[12spdf]$} & \multicolumn{1}{c}{$[20s19pdfg]$} & \multicolumn{1}{c}{} \\
  State        & Polariz.   &         & 11$e$ CI  & 17$e$ CI  & 11$e$ CI & Final           \\
\hline \\ [-0.6pc]
$^2\!D_{5/2}$  & $\alpha_0$ &  Static &    93     &    95     &   84     &  95   \\[0.2pc]
               &            &  532 nm &    65     &    65     &   58     &  65   \\[0.2pc]
               &            & 1064 nm &    56     &    56     &   49     &  56   \\[0.5pc]

               & $\alpha_2$ &  Static &   -40     &   -41     &   -34    & -41   \\[0.2pc]
               &            &  532 nm &   -0.8    &   -2.1    &   -0.7   &  -2  \\[0.2pc]
               &            & 1064 nm &   -1.0    &   -0.6    &   -1.0   &  -1  \\[0.5pc]

$^2\!D_{3/2}$  & $\alpha_0$ &  Static &    53     &    53     &   46     &  53   \\[0.2pc]
               &            &  532 nm &    62     &    65     &   59     &  65   \\[0.2pc]
               &            & 1064 nm &    57     &    56     &   50     &  56   \\[0.5pc]

               & $\alpha_2$ &  Static &    0.6    &    1.5    &   2.0    &  1.5  \\[0.2pc]
               &            &  532 nm &   -0.07   &   -0.9    &   10     &  -1  \\[0.2pc]
               &            & 1064 nm &   -1.2    &   -0.8    &    6     &  -1
\end{tabular}
\end{ruledtabular}
\end{table}
% ####################################################################################################################

As seen in~\tref{Tab:PolarCI}, the values of the scalar polarizabilities $\alpha_0$ obtained within the framework of 11$e$ and 17$e$ CI
are practically the same. This means that they are insensitive to the addition of the $4p$ electrons to the valence field.
Tensor polarizabilities are small in all cases, except static $\alpha_2(^2\!D_{5/2})$. Its relatively large value
is determined by the contribution of the intermediate state $4d^{10}5p \,\, ^2\!P_{3/2}$ separated from
the $^2\!D_{5/2}$ state by a small energy interval of 230 cm$^{-1}$.
We consider the results obtained within the framework of the 17$e$ CI as final.

Our static scalar polarizability $\alpha_0(^2\!D_{5/2}) = 95\,\, {\rm a.u.}$ differs by two times from the value
$47(2)\,\, {\rm a.u.}$ obtained in Ref.~\cite{DzuAllFla21}. As mentioned in Ref.~\cite{DzuAllFla21}, the summation
over intermediate states in \eref{alpha} is strongly dominated by the states of the $4d^9 5s5p$ configuration.
This is true for all the cases considered, except for the static scalar and tensor polarizabilities of the $^2\!D_{5/2}$ state.
Due to the small energy difference between the $^2\!D_{5/2}$ and $4d^{10}5p \,\, ^2\!P_{3/2}$ states,
the latter gives a contribution of 40\% to $\alpha_0(^2\!D_{5/2})$.
If this contribution was missed in Ref.~\cite{DzuAllFla21}, it could explain this discrepancy.

To correctly account for the contribution of the intermediate state $4d^{10}5p \,\, ^2\!P_{3/2}$ in \eref{alpha}, we used
the experimental energy difference $E(4d^{10}5p \,\,^2\!P_{3/2}) - E(^2\!D_{5/2})$.
As seen from \tref{E:CI}, there is good agreement between the theoretical and experimental energies for the high-lying
states of the $4d^9 5s5p$ configuration. Replacing theoretical energies with experimental ones in contributions
of these terms had virtually no effect on the values of polarizabilities.

We present in \tref{Tab:ME} the dominant contributions of individual odd-parity states to the scalar static $^2\!D_{5/2}$ and $^2\!D_{3/2}$ polarizabilities calculated in the framework of the 17$e$ CI method.
These contributions are listed separately in the column labeled ``$\alpha_0$'', with the corresponding absolute values of the reduced
electric-dipole matrix elements given in the column labeled ``$D$'' (in a.u.). The experimental~\cite{RalKraRea11}
transition energies are given in column $\Delta E$ (in cm$^{-1}$). The remaining contributions to the polarizabilities are
given in rows labeled ``Other.''

We note that the contribution of the $4d^{10}5p \,\, ^2\!P_{3/2}$ state to the scalar
static polarizability of the $^2\!D_{3/2}$ state is very small, in contrast to $\alpha_0(^2\!D_{5/2})$.
This is because the energy difference $E(4d^{10}5p \,\,^2\!P_{3/2}) - E(^2\!D_{5/2})$ is 18 times smaller than
$E(4d^{10}5p \,\,^2\!P_{3/2}) - E(^2\!D_{3/2})$, while the reduced matrix element $\langle ^2\!D_{5/2} || d || ^2\!P_{3/2} \rangle$
is 3 times larger than $\langle ^2\!D_{3/2} || d || ^2\!P_{3/2} \rangle$.
% ###################################################################################################################
\begin{table}[t]
\caption{The contribution of individual states to the static scalar polarizabilities $\alpha_0(^2\!D_{5/2})$ and
$\alpha_0(^2\!D_{3/2})$ (in a.u.) are presented. The dominant contributions to the polarizabilities are listed separately
in the column labeled ``$\alpha_0$'' with the corresponding absolute values of electric-dipole reduced matrix elements given
in the column labeled ``$D$'' (in a.u.). The experimental~\cite{RalKraRea11} transition energies are given in column
$\Delta E$ (in cm$^{-1}$). The remaining contributions to the polarizabilities are given in rows labeled ``Other.''}
\label{Tab:ME}%
\begin{ruledtabular}
\begin{tabular}{ccccc}
    State      &              Contribution               & $\Delta E$ &  $D$  & $\alpha_0$  \\
\hline \\ [-0.6pc]
$^2\!D_{5/2}$  & $^2\!D_{5/2} -\,4d^{10}5p\,\,^2\!P_{3/2}$ &     230    &  0.6  &    40        \\[0.2pc]
               & $^2\!D_{5/2} -\,4d^9 5s5p\,\,^2\!P_{3/2}$ &   41942    &  4.2  &    10        \\[0.2pc]
               & $^2\!D_{5/2} -\,4d^9 5s5p\,\,^2\!F_{7/2}$ &   42092    &  6.0  &    21        \\[0.2pc]
               & $^2\!D_{5/2} -\,4d^9 5s5p\,\,^2\!D_{5/2}$ &   43285    &  4.9  &    14        \\[0.2pc]
               &              Other                        &            &       &    10        \\[0.2pc]
               &              Total                        &            &       &    95        \\[0.5pc]
$^2\!D_{3/2}$  & $^2\!D_{5/2} -\,4d^9 5s5p\,\,^2\!P_{1/2}$ &   46162    &  3.0  &     8        \\[0.2pc]
               & $^2\!D_{5/2} -\,4d^9 5s5p\,\,^2\!F_{5/2}$ &   46568    &  5.1  &    22        \\[0.2pc]
               & $^2\!D_{5/2} -\,4d^9 5s5p\,\,^2\!D_{3/2}$ &   47700    &  4.1  &    14        \\[0.2pc]
               &              Other                        &            &       &     7        \\[0.2pc]
               &              Total                        &            &       &    53
\end{tabular}
\end{ruledtabular}
\end{table}
% ####################################################################################################################

It is problematic to determine the exact values of the uncertainties of these polarizabilities. We can estimate the quality of the wave functions of the 
$4d^9 5s^2\,\,^2\!D_{5/2,3/2}$ states by comparing their lifetimes $\tau(^2\!D_{5/2}) = 0.15\, {\rm s}$ and $\tau(^2\!D_{3/2}) = 68\, \mu{\rm s}$ found in this work with the experimental results, 0.2 s~\cite{Gar64} and 40 $\mu$s~\cite{BadPliJun04}, respectively.
Taking into account that the uncertainties are not assigned to the experimental values, we can assume that
they can be 100\% or even more. On the basis of this, the agreement between the theory and experiment seems reasonable.

A comparison of the results for polarizabilities obtained in the framework of 11$e$ and 17$e$ CI is presented in
\tref{Tab:PolarCI}. Here, we see that both the scalar and the tensor polarizabilities are rather insensitive to the core-valence correlations. As follows from the comparison of the
11$e$ $[12spdf]$ and $[20s19spdfg]$ CI calculations (see \tref{Tab:ME}), a sensitivity of the polarizabilities to the valence-valence correlations is larger. Based on this difference, we estimate the uncertainties of the scalar polarizabilities and the static tensor $^2\!D_{5/2}$ polarizability at the level of 15-20\%. Other tensor polarizabilities are small.
This is due to large cancellations between the main contributions. We consider these values to be order-of-magnitude estimates.

\section{Conclusions}
To conclude, we carried out calculations of the energies, $E1$ transition amplitudes, and static and dynamic
polarizabilities of the low-lying states, including the states with the unfilled $4d$ shell $4d^9 5s5p \,\,^2\!D_{5/2,3/2}$.
To study the properties of the states belonging to the configuration $4d^{10} x$ (where $x \equiv 5,6s; 5,6p; 5d$), we used the
single-electron approaches combining DHF with MBPT and the all-order method. By comparing the results obtained within
the framework of these two methods, we assign uncertainties to the values obtained.

The properties of the $4d^9 5s5p \,\,^2\!D_{5/2,3/2}$ states were studied within the framework of the 11$e$ and 17$e$ CI methods.
We carried out analyses of the different contributions to the dc and ac $^2\!D_{5/2,3/2}$  polarizabilities
and determined the odd-parity states that gave the main contribution. The electric-dipole transition amplitudes from these states
to the $^2\!D_{5/2,3/2}$ states were determined and discussed.
%========================
\section{Acknowledgments}
%========================
We thank David DeMille and Zoe Yan for bringing this problem to our attention and for useful discussions.
This work is part of the ``Thorium Nuclear Clock'' project that has received funding from the European Research Council under the European Union's Horizon 2020 research and innovation program (Grant No. 856415).  The calculations in this work were done through the use of Information Technologies resources at the University of Delaware, specifically the high-performance Caviness and DARWIN computer clusters.  This work was supported by National Science Foundation Grant No.~PHY-2309254, and Office of Naval Research Grant No. N000142512105

%\bibliography{Ag}


\end{document}